%% file: main.tex
\documentclass[pdftex,11pt,oneside,letter]{article}

\usepackage[utf8]{inputenc}
\usepackage[T1]{fontenc}
\usepackage[english]{babel}
\DeclareUnicodeCharacter{03A6}{\ensuremath{\Phi}}

\usepackage{amsmath,amsfonts,amssymb,amscd}
\usepackage{mathtools}
\usepackage{siunitx}
\usepackage{steinmetz}
\usepackage{gensymb}

\usepackage{graphicx}           
\usepackage{tabularx}
\usepackage{booktabs}
\usepackage{tikz}
\usepackage[font=footnotesize,labelfont=bf]{caption}
\usepackage{subcaption}
\usepackage[hidelinks]{hyperref}

\usepackage{orcidlink}

\usepackage{pgfplots}
\usepgfplotslibrary{groupplots}
\usepackage{paralist}
\usepackage{enumitem}
\usepackage[export]{adjustbox}
\usepackage{multicol}
\usepackage{wrapfig}
\usepackage{titlesec}
\usepackage[left=1in,top=1in,right=1in,bottom=1in]{geometry}
\usepackage{setspace}

\usepackage{booktabs,tabularx,ragged2e}
\newcolumntype{L}[1]{>{\RaggedRight\arraybackslash}p{#1}}
\newcolumntype{Y}{>{\RaggedRight\arraybackslash}X}

\usepackage{fancyhdr}
\usepackage[numbers,sort&compress]{natbib}

\usepackage[hidelinks]{hyperref}

\usepackage{lipsum}

\setdescription{leftmargin=0.5\parindent,labelindent=\parindent}

\graphicspath{{./figures/}}

\titlespacing{\section}{0pt}{*2}{*0}
\titlespacing{\subsection}{0pt}{*1}{*0}
\titlespacing{\subsubsection}{0pt}{*0}{*0}

\let\OLDthebibliography\thebibliography
\makeatletter
\renewcommand\thebibliography[1]{%
  \OLDthebibliography{#1}%
  \setlength{\parskip}{0pt}%
  \setlength{\itemsep}{0pt plus 0.3ex}%
}
\makeatother

\usepackage[nolist,nohyperlinks]{acronym}

\begin{document}

\title{\large\bfseries\vspace{-8ex}%
Konzepte zur Effizienzsteigerung von Traktionsmotoren in batterieelektrischen Fahrzeugen durch den Einsatz neuartiger
teillastoptimierbarer Motor- und Invertertopologien
  \vspace{2ex}}
\author{\normalsize Christoph Sachs\,\orcidlink{0009-0005-5559-9342}$^{1,2}$, 
Martin Neuburger$^{2}$\\[0.8ex]
\small $^{1}$Universität Stuttgart, Institut für Systemdynamik, Waldburgstraße 19, 70563 Stuttgart, Deutschland\\
\small $^{2}$Hochschule Esslingen, Campus Göppingen, Robert-Bosch-Straße 1, 73037 Göppingen, Deutschland}

\date{}
\maketitle

\fancyhf{}                
\fancyfoot[C]{\thepage}   

\setcounter{page}{151} 
\pagestyle{fancy}

\renewcommand{\figurename}{Fig.}

\input{abstract.tex}
\input{introduction.tex}

\input{section1.tex}
\input{section2.tex}
\input{section3.tex}

\input{conclusion.tex}

\input{99_acronymsA.tex}

{\setstretch{1}\vspace{\baselineskip}%
\bibliographystyle{IEEEtran}
\bibliography{bibligraphy}}

\end{document}

%% file: abstract.tex
\textbf{Abstract}: 

Um die Effizienz zukünftiger Elektrofahrzeuge zu erhöhen, ist es entscheidend, die Verluste im Antriebsstrang von batteriebetriebenen Fahrzeugen zu reduzieren. Dies ermöglicht die Erhöhung der Reichweite oder eine Gesamtkostenersparnis durch Reduktion der Batteriekapazität bei gleichbleibender Reichweite. Harmonische Motorverluste machen einen vermeidbaren Verlustanteil des kompletten eDrives von über 30\% in Standard-B6-2L 300\,kW iPMSM Konfigurationen aus. Diese Verluste resultieren aus einer hochfrequenten Spannungsverzerrung über den Motorwindungen, die durch verschiedene Ansätze reduziert werden können. Von großer Bedeutung ist hierbei die Einordnung kostenneutraler und preiswerter Konzepte zur Verlustreduktion. Folgend werden Ansätze zur Verlustreduktion vorgestellt und eingeordnet, die innerhalb der letzten Jahre seitens der Forschung und Industrie erarbeitet wurden. Explizit werden neuartige teillastfähige Motor- und Inverterkonzepte vorgestellt, die eine Motorumschaltung oder einen Mehrlevel-Betrieb zur Reduktion der harmonischen Verluste im Teillastbereich bewirken. 

%% file: introduction.tex
\section{Einführung}
Die Effizienz des Inverters spielt eine zentrale Rolle für die Leistung und Kosten des Antriebsstrangs in Elektrofahrzeugen \cite{sachs2025modloss}. Der Inverter wandelt den Gleichstrom der Batterie in den Wechselstrom um, der den Elektromotor antreibt, und steuert präzise die Leistungsabgabe. Eine hohe Effizienz des Inverters minimiert Energieverluste und erhöht die Reichweite des Fahrzeugs. Gleichzeitig beeinflussen die verwendeten Halbleitermaterialien die Kosten: Moderne Materialien wie Siliziumkarbid ermöglichen geringere Schaltverluste und tragen so zur Effizienzsteigerung bei, erhöhen jedoch durch die größere Chipflächen die Gesamtkosten. Die richtige Balance zwischen Effizienz und Kosten ist daher entscheidend. 

\begin{figure}
    \centering
    \includegraphics[width=0.7\linewidth]{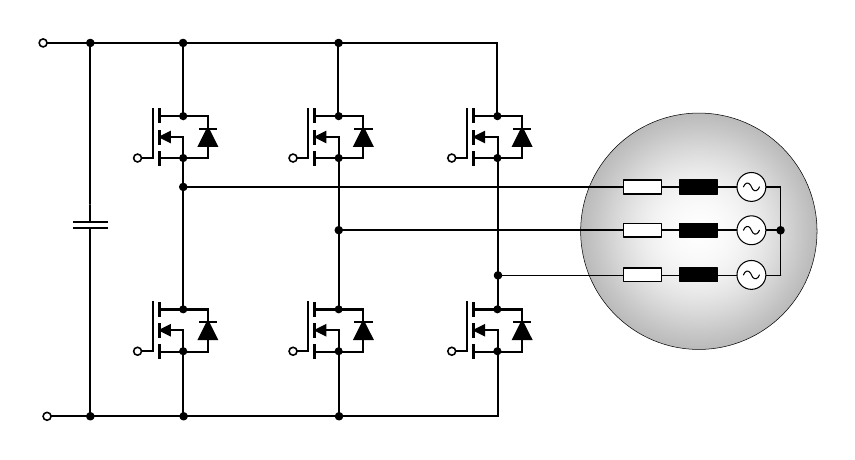}
    \caption{B6-2L SiC MOSFET Inverter an Zwi-
schenkreiskondensator und 3-Phasen-Motor in Sternschaltung angebunden.}
    \label{fig:2L-B6}
\end{figure}

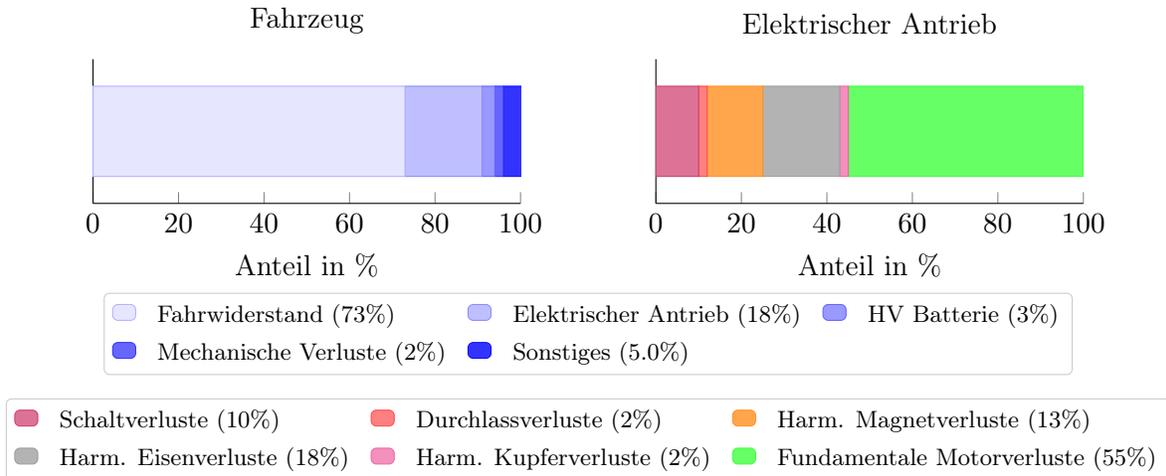
\begin{figure}[htbp]
\centering
\input{tikz/A/losses_overview}
\caption{Gemittelte Verlustanteile eines Batterieelektrischen Fahrzeugs links \cite{RambetiusCTI2023} und eines eDrives mit SiC MOSFET aufgebauten B6-2L Inverter und iPMSM rechts, betrieben mit 10\,kHz SVPWM an einem 300\,kW Traktionsmotor über den WLTP-Fahrzyklus \cite{SachsECCE2024}.}
\label{fig:2}
\end{figure}

Abbildung 2 stellt die Verlustanteile eines 2-Level-Inverters und eines iPMSM (interior Permanent Magnet Synchronous Motors) über den WLTP (Worldwide Harmonized Light Vehicles Test Procedure) dar. Die Inverterverluste setzen sich aus Schaltverlusten (10 \%) und Durchlassverlusten (2 \%) zusammen. Die harmonischen Verluste in der Maschine, bestehend aus magnetischen Verlusten (13 \%), Eisenverlusten (18 \%) und Kupferverlusten (2 \%), tragen einen signifikanten Anteil von insgesamt 33 \% zu den Verlusten bei. Der größte Verlustanteil entfällt jedoch auf die fundamentalen Motorverluste, welche mit 55 \% die dominierende Verlustquelle darstellen \cite{SachsECCE2024}. Die fundamentalen Motorverluste stellen den Verlustanteil dar, welcher die elektrischen Verluste aufgrund der Motordrehfrequenz entstehen. Hierbei wird meistens zwischen Kupferverlusten und Eisenverlusten unterschieden, da die Magnetverluste bei der niedrigen elektrischen Motordrehfrequenz zu vernachlässigen sind. Die harmonischen Motorverluste hingegen entstehen aufgrund der Spannungsverzerrung über den Motorwicklungen durch das hochfrequente Schalten der Halbleiterschalter, welche im Wechselrichter eingesetzt werden.
Auf der linken Seite der Abbildung 2 sind die gesamten Fahrzeugverluste in ihre einzelnen Anteile aufgeschlüsselt. Den größten Beitrag leistet der Fahrwiderstand, der durch aerodynamische Einflüsse, Rollwiderstände und andere äußere Kräfte entsteht, die das Fahrzeug überwinden muss. Einen weiteren erheblichen Anteil tragen die Verluste des elektrischen Antriebs bei, die aus Umwandlungs- und Wirkungsgradverlusten resultieren. Die Hochvolt-Batterie verursacht ebenfalls Verluste, die vor allem durch ihren Innenwiderstand sowie Lade- und Entladeprozesse bedingt sind. Verluste im Getriebe entstehen hauptsächlich durch Reibung und mechanische Ineffizienzen, wobei ihr Anteil an den Gesamtverlusten vergleichsweise geringer ausfällt. Ergänzt wird die Aufteilung durch die Verluste, die von Hilfsverbrauchern wie Klimaanlage, Beleuchtung oder anderen elektrischen Geräten verursacht werden. Diese Darstellung verdeutlicht, welche Komponenten in einem batterieelektrischen Fahrzeug über den WLTP-Fahrzyklus die größten Energieverluste aufweisen \cite{RambetiusCTI2023}.

%% file: tikz/A/losses_overview.tex
\pgfmathsetmacro{\Fahr}{73}
\pgfmathsetmacro{\Getr}{2}
\pgfmathsetmacro{\EDrive}{18}
\pgfmathsetmacro{\HV}{3}
\pgfmathsetmacro{\Sonst}{max(5, 100 - (\Fahr+\Getr+\EDrive+\HV))}

\begin{tikzpicture}
\begin{groupplot}[
  group style={group size=2 by 1, horizontal sep=18mm},
  xbar stacked,
  xmin=0, xmax=100,
  width=0.44\linewidth,
  height=35mm,
  /pgf/bar width=12mm,
  xlabel={Anteil in \%},
  ytick=data, yticklabels={},
  axis x line*=bottom,
  axis y line*=left,
  enlarge y limits=0.25,
]

\nextgroupplot[
  title={Fahrzeug},
  legend to name=AggLegend,      
  legend columns=3,
  legend style={
    draw=black!20, fill=white, rounded corners=2pt,
    font=\footnotesize, column sep=6pt, row sep=2pt,
    legend cell align=left,
  },
]
\addplot+[fill=blue!10,draw=blue!30] coordinates {(\Fahr,0)}; 
\addplot+[fill=blue!25,draw=blue!45] coordinates {(\EDrive,0)}; 
\addplot+[fill=blue!40,draw=blue!60] coordinates {(\HV,0)};    
\addplot+[fill=blue!60,draw=blue!80] coordinates {(\Getr,0)};  
\addplot+[fill=blue!80,draw=blue!95] coordinates {(\Sonst,0)}; 
\legend{Fahrwiderstand (\Fahr\%), Elektrischer Antrieb (\EDrive\%), HV Batterie (\HV\%), Mechanische Verluste (\Getr\%), Sonstiges (\Sonst\%)}

\nextgroupplot[
  title={Elektrischer Antrieb},
  legend to name=DetLegend,
  legend columns=3,
  legend style={
    draw=black!20, fill=white, rounded corners=2pt,
    font=\footnotesize, column sep=6pt, row sep=2pt,
    legend cell align=left,
  },
]
\addplot+[fill=purple!55,draw=purple!75]   coordinates {(10,0)}; 
\addplot+[fill=red!50,draw=red!70]         coordinates {(2,0)};
\addplot+[fill=orange!70,draw=orange!85]   coordinates {(13,0)};
\addplot+[fill=gray!60,draw=gray!75]       coordinates {(18,0)};
\addplot+[fill=magenta!55,draw=magenta!75] coordinates {(2,0)};
\addplot+[fill=green!60,draw=green!75]     coordinates {(55,0)};
\legend{Schaltverluste (10\%), Durchlassverluste (2\%), Harm. Magnetverluste (13\%), Harm. Eisenverluste (18\%), Harm. Kupferverluste (2\%), Fundamentale Motorverluste (55\%)}

\end{groupplot}

\path (group c1r1.south east) -- (group c2r1.south west) coordinate[pos=0.5] (groupbottom);
\node[anchor=north] at ($(groupbottom)+(0,-30pt)$) {\pgfplotslegendfromname{AggLegend}};
\node[anchor=north] at ($(groupbottom)+(0,-70pt)$) {\pgfplotslegendfromname{DetLegend}};

\end{tikzpicture}

%% file: section1.tex
\section{Harmonische Motorverluste}
Entsprechend Abbildung 2 stellen die harmonischen Motorverluste einen wesentlichen und vermeidbaren Anteil der gesamten elektrischen Antriebsstrangverluste dar. Entsprechend \cite{SachsECCE2024} wird die harmonische Spannungswelligkeit an den Motorwicklungen in d- und q-Achsen-Komponenten  und  aufgeteilt und mittels Fast Fourier Transformation (FFT) in das Frequenzspektrum transformiert, was die Identifizierung einzelner harmonischer Komponenten ermöglicht. Diese Oberschwingungen werden mit spezifischen harmonischen Motorparametern verrechnet, einschließlich der Motorinduktivitäten  und  der d- und q-Achse. Es werden harmonische Eisen-, Magnet- und Kupferverlustfaktoren ,  und  verwendet, wobei die ersten beiden nicht direkt ohmsche Widerstände darstellen. Stattdessen geben sie einen äquivalenten ohmschen Verlust aufgrund von Wirbelströmen an. Die Skalierungsfaktoren , und  werden auf der Grundlage eines analytischen Modells verwendet, das an die Motormessdaten angepasst wurde. Es ist wichtig, zu beachten, dass diese Variablen undParameter je nach der verwendeten Motorkonfiguration und den Ziellastpunkten unterschiedlich sind. Für die betrachteten Frequenzkomponenten wurden die Minimal- und Maximalfrequenzen  und  gewählt, da hierdurch nahezu alle harmonischen Verlustanteile berücksichtigt werden. \cite{sachs2025modloss}
Die Kupferverluste können entsprechend auftretender Skin- und Proximity-Effekten der Motorwindungen über folgende Gleichung ermittelt:

\begin{equation}
\label{eq:pharmcu}
P_{cu,h} = k_{{cu}} \cdot \sum_{f_{{min}}}^{f_{{max}}} \left[ \frac{R_{\text{cu},h}}{f_h^2} \cdot \left( \frac{U_{d,h}^2}{L_{d,h}^2} + \frac{U_{q,h}^2}{L_{q,h}^2} \right) \right]
\end{equation}

Die Eisenverluste und Magnetverluste basieren auf hochfrequenten Hyserese-, Wirbelstrom- und Excess-Verlusten \cite{RambetiusCTI2023,SchweizerTIE2012,MarmolejoKIT2024} die durch folgende Gleichungen für die Eisenverluste

\begin{equation}
\label{eq:pharmiron}
P_{{iron},h} = k_{{iron}} \cdot \sum_{f_{{min}}}^{f_{{max}}} \left[ \frac{U_{d,h}^2+U_{q,h}^2}{R_{\text{iron},h}} \right]
\end{equation}

und für die Magnetverluste berechnet werden können

\begin{equation}
\label{eq:pharmmagnet}
P_{{mag},h} = k_{{mag}} \cdot \sum_{f_{{min}}}^{f_{{max}}} \left[ \frac{U_{d,h}^2}{R_{\text{mag},h}} \right]
\end{equation}

Diese Berechnungsvarianten stellen ein Optimum aus Rechenaufwand und Genauigkeit der Ergebnisse für die Zielfunktionen dar. Alternative Möglichkeiten werden unter anderem in \cite{WeberSpringer2024} erläutert. Entsprechend Gleichung (1), (2) und (3) können verschiedene Maßnahmen ergriffen werden, um die harmonischen Verlustkomponenten zu reduzieren. Dies sind beispielsweise konstruktive Motoranpassungen, sodass die Verlustfaktoren , und  angepasst werden, um die resultierenden Verluste zu reduzieren. Dies kann exemplarisch durch dünnere Motorbleche für die Reduktion der harmonischen Eisenverluste durch geringere Wirbelstromeffekte, den Einsatz von segmentierten Magneten zur Reduktion von Magnetverlusten oder die Nutzung von Hairpin-Windungen, um den Einfluss des Proximity-Effekts auf die harmonischen Kupferverluste zu reduzieren. Ein weiterer Aspekt ist die Betrachtung welche Arbeitspunkte besonders relevant für die Gesamteffizienz des Antriebsstrangs sind. Hierdurch können in diesem Teillastbereich Ansteuerungsstrategien durch optimierte Hardware oder Software angewandt werden, um die Gesamtverluste zu reduzieren. Folgend werden verschiedene Konzepte betrachtet, welche zu einer Effizienzsteigerung durch neue Konzepte zur harmonischen Verlustreduktion führen.

%% file: section2.tex
\section{Konzepte}
Die dargestellten Topologien verdeutlichen die Bandbreite der Ansätze zur Realisierung von Traktionsinvertern. Jede der vorgestellten Varianten erfüllt spezifische Anforderungen, wie Effizienz, Lastflexibilität oder Leistungsfähigkeit, und zeigt, wie moderne Halbleitertechnologien wie SiC-MOSFETs für optimale Leistung und Betriebseffizienz eingesetzt werden können.

\subsection{Volllastfähige teillastoptimierbare Inverter-/Motortopologien}

Ein volllastfähiges und teillastoptimierbares System ist ein technisches System, dessen Komponenten und Betriebsmodi so ausgelegt werden können, dass eine maximale Effizienz im Teillastbereich erreicht wird. Dies wird durch eine gezielte Ansteuerungsstrategie ermöglicht, die zwischen Betriebsmodi für Teillast und Volllast wechselt. Im Teillastmodus werden spezifische Systemkomponenten eingesetzt oder anders betrieben, um Verluste zu minimieren und die Systemeffizienz zu steigern. Im Volllastmodus werden hingegen andere Komponenten genutzt, um hohe Leistungsanforderungen zu erfüllen. Ein solches Design erlaubt es, kritische Komponenten gezielt für den Teillastbetrieb zu optimieren, z. B. durch Kostenreduktion bei Leistungshalbleitern, indem kleinere Chips genutzt werden. Teillastoptimierbare Inverter-/Motortopologien stellen sehr wichtige Konzepte für batterieelektrische Fahrzeuge da, denn sie ermöglichen einen neuen Lösungsraum für Effizienz/Kosten-optimale Antriebsstränge \cite{SachsECCE2024}.

\subsection{Teillastoptimierbare Mehrlevel-Inverter}
\begin{figure}
    \centering
    \includegraphics[width=0.7\linewidth]{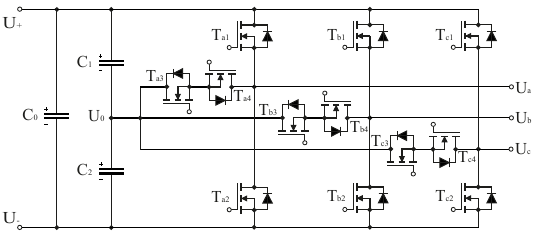}
    \caption{TNPC-3L SiC MOSFET Inverter an Zwischenkreiskondensator und 3-Phasen-Motor in Sternschaltung angebunden.}
    \label{fig:TNPC}
\end{figure}

Abbildung 3 zeigt einen TNPC-3L SiC-MOSFET-Inverter. Dieser T-Type Neutral Point Clamped (TNPC) Inverter nutzt ebenfalls SiC-MOSFETs, bietet jedoch im Vergleich zur 2-Level-Konfiguration eine verbesserte Spannungsverteilung und geringere Schaltverluste. Durch die 3-Level-Topologie wird eine bessere Effizienz, insbesondere im Teillastbetrieb, erreicht. Wird die Topologie nur mit einer Chipfläche der Mittelpunktschalter zur Erfüllung des Teillastbereichs ausgestattet, so kann ein Effizienz-Kosten-Design-Optimum erreicht werden \cite{SachsECCE2024}. Eine weitere Kostenreduktionsmöglichkeit bieten teillastoptimierte Zwischenkreise, wie in Abbildung \ref{fig:TNPC} dargestellt \cite{sachsASIA2025}.

\subsection{Teillastoptimierbarer Tiefsetzsteller mit 2-Level-Inverter}
Die Verringerung der Spannungripple ist neben der Erhöhung der Spannungslevelanzahl auch über eine dynamische Reduktion der Zwischenkreisspannung durch einen Buck-Converter realisierbar. Abbildung 5 stellt eine Variation des klassischen 2-Level-Inverters dar. Der B6-2L SiC-MOSFET-Inverter in dieser Konfiguration ist mit einem integrierten teillastfähigen Tiefsetzsteller ausgestattet. Dieser Tiefsetzsteller dient dazu, die Zwischenkreisspannung dynamisch an die Betriebsbedingungen anzupassen, wodurch der Energieverbrauch im Teillastbetrieb reduziert wird. Trotz der zusätzlichen Komplexität durch den DC/DC-Wandler bleibt die Grundstruktur des Inverters unverändert, was die Integration in bestehende Systeme erleichtert und den Betrieb effizienter gestaltet. Entsprechend \cite{KimPatent2024} ist bei der Auslegung des DC/DC-Wandlers besonders auf einen kosten-effizientes Design zu achten, sodass neben der Drosselinduktivität auch die zusätzlichen Halbleiterschalter zu einem optimierten Kosten/Effizienz-Metrik des Gesamtsystems führen. Auch alternative DC/DC-Wandler können hierbei berücksichtigt werden. \cite{KimPatent2024} vergleicht exemplarischen einen Cuk-Converter mit einem klassischen Tiefsetzsteller, wobei der Cuk-Converter keiner wesentlichen Effizienz/Kosten-Vorteile in Kombination mit einem B6-Inverter ermöglicht.

\begin{figure}
    \centering
    \includegraphics[width=0.7\linewidth]{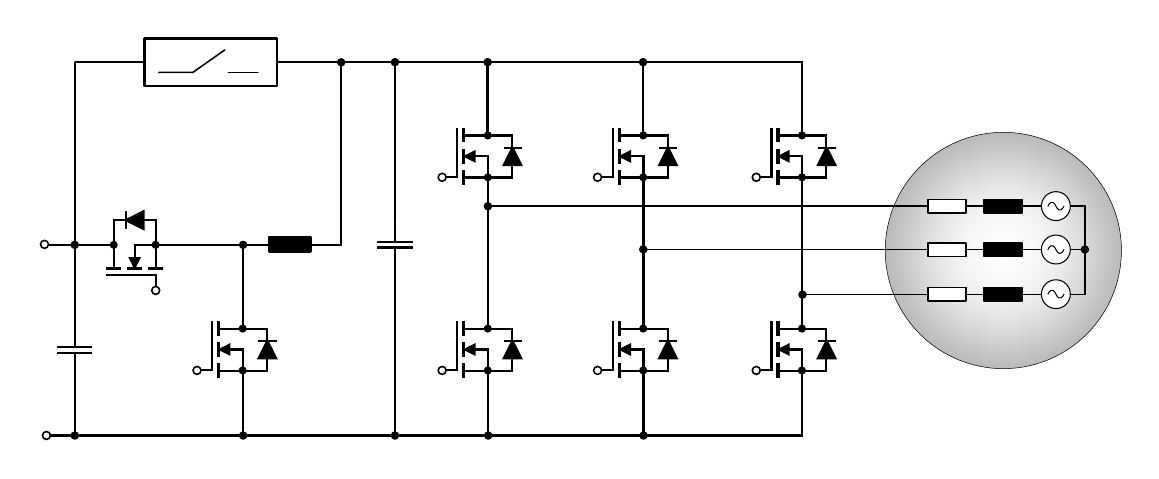}
    \caption{B6-2L SiC MOSFET Inverter mit teillastfähigemTiefsetzsteller \cite{VelicOJPE2023}.}
    \label{fig:DCDC}
\end{figure}

\begin{figure}
    \centering
    \includegraphics[width=1\linewidth]{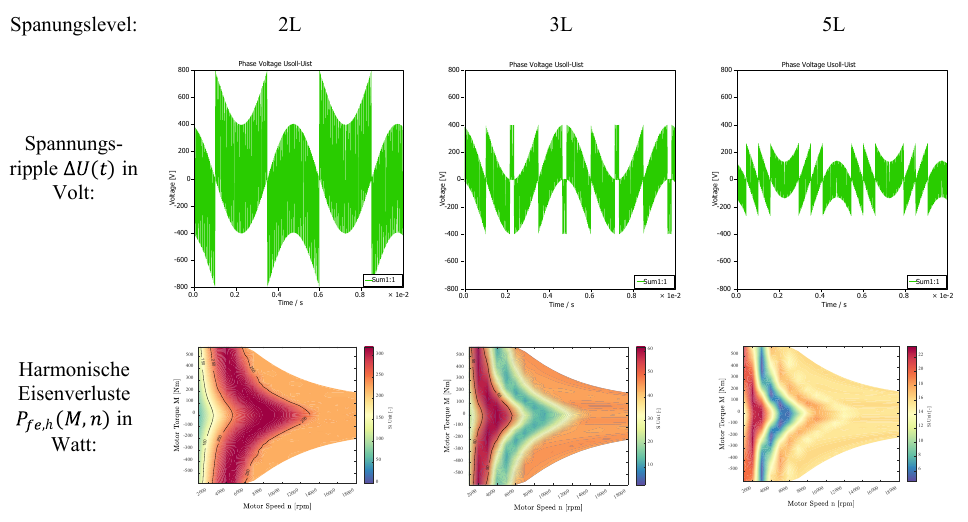}
    \caption{Spannungslevel (2L, 3L, 5L) und deren betriebspunktabhängigen Reduktion der harmonischen Eisen-Motorverluste einer 300kW iPMSM.}
    \label{fig:2L3L5L}
\end{figure}

\subsection{Teillastoptimierbare Inverter mit Fähigkeit zur Motorumschaltung}
Abbildung 6 zeigt die Topologie eines B6²-Y Inverters, bei dem der Motor mit offenen Wicklungen (Open-Winding) verbunden ist. Diese Konfiguration ermöglicht eine flexiblere Ansteuerung des Motors und erlaubt eine Motorumschaltung zum Betrieb des Motors in Sternschaltung. Der Betrieb mit offenen Windungen wird durch die linke und rechte B6-Inverterhälften ermöglicht. Hier kann der Inverter als B6²-Konfiguration betrieben werden. Durch Aktivierung der zusätzlichen teillastoptimierbaren IGBTs wird ein Betrieb des Motors im Sternpunkt ermöglicht. Dies ermöglicht die applizierte Spannung um  reduziert. Dadurch ergeben sich reduzierte harmonische Motorverluste entsprechend Abbildung 7. Zu beachten ist hierbei auch der reduzierte Betriebsbereich, da nicht alle stellbaren Drehmomente im Feldschwächbereich erreicht werden können, da nicht der erforderliche Strom für diese Betriebspunkte in die Motorwindungen mit der reduzierten Spannung eingespeist werden kann.

\begin{figure}
    \centering
    \includegraphics[width=1\linewidth]{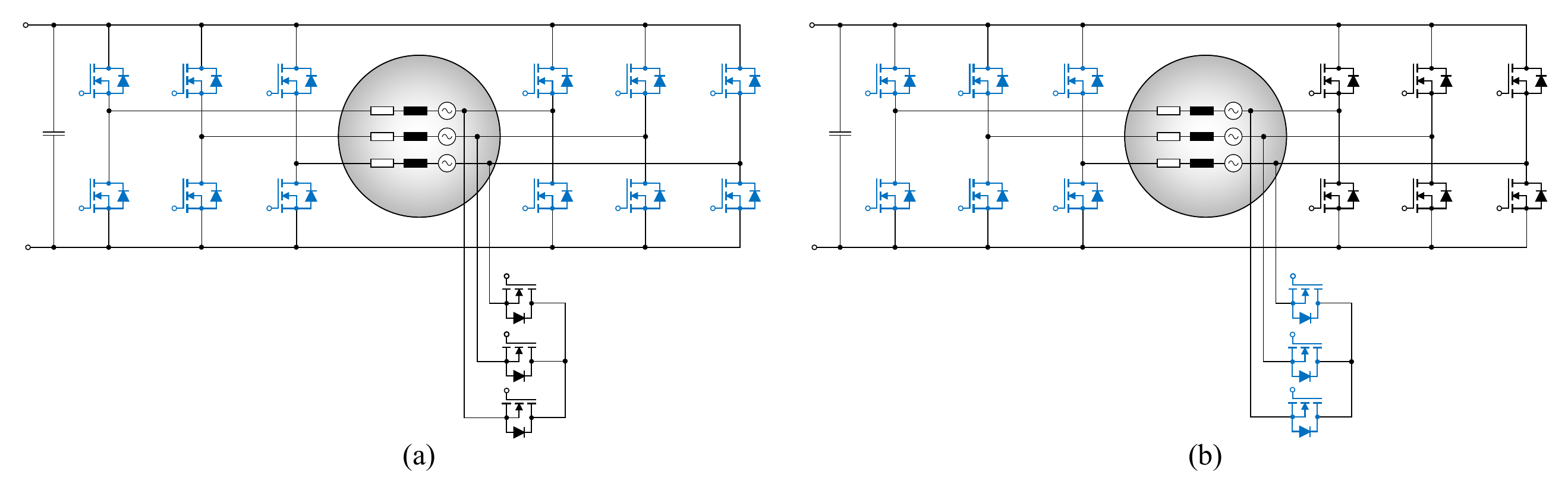}
    \caption{B6$^2$-Y inverter an Zwischenkreiskondensator und Motor mit offenen Windungen angebunden für (a) H-Mode, (b) Y-Mode.}
    \label{fig:B62Y}
\end{figure}

\begin{figure}
    \centering
    \includegraphics[width=1\linewidth]{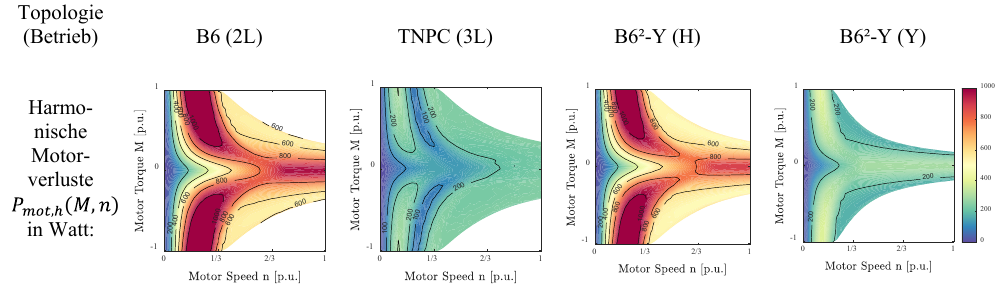}
    \caption{Gesamte Harmonische Motorverluste von verschiedenen Topologien und Betriebsstrategien einer 300kW iPMSM.}
    \label{fig:harm}
\end{figure}

%% file: section3.tex
\subsection{Sonstiges}

Um die Kosten von Invertern weiter zu reduzieren und die Effizienz des Antriebsstrangs zu erhöhen, gibt es verschiedene Ansätze, die auf technischer und systemischer Ebene umgesetzt werden können. Eine Möglichkeit besteht in der Anpassung der Schaltfrequenz \cite{GeestICEM2014}, um das Optimum zwischen Systemverlusten und anderen Anforderungen wie dem zulässigen Spannungsripple über den Motorwindungen zu optimieren. Die Variation der eingesetzten Schalter, beispielsweise durch die Kombination von Wide-Bandgap-(WBG)-Halbleitern mit konventionellen Schaltern, erlaubt es, die Effizienz zu maximieren und gleichzeitig die Systemkosten durch Batterieverkleinerung zu senken. Dieser Ansatz wird als Schalterhybridisierung bezeichnet und ist insbesondere für Anwendungen mit variablen Lastanforderungen interessant.
Zusätzlich kann durch die Einführung von Fahrzeugkonfigurationen mit mehreren Motoren eine bessere Anpassung der Antriebsleistung an die Fahrsituation erreicht werden durch optimierte Ansteuerverfahren. Dies erhöht die Effizienz des Gesamtsystems und reduziert Verluste im Teillastbereich. Eine weitere Kosten- und Effizienzoptimierung wird durch die ganzheitliche (holistische) Betrachtung der Komponenten erreicht, etwa durch die Optimierung der Zwischenkreiskondensatoren \cite{SachsECCEASIA2025_SplitCap}. Dabei werden nicht nur die elektrischen Eigenschaften, sondern auch thermische und mechanische Aspekte berücksichtigt. Ergänzend dazu ermöglicht eine holistische Softwarestrategie, die sowohl die Effizienz als auch die Lebensdauer der Systeme optimiert, durch intelligente Regelungsansätze eine weitere Steigerung der Gesamtleistung \cite{SachsDamageAvoidance2024}.
Smarte Gate-Treiber spielen ebenfalls eine zentrale Rolle bei der Effizienzsteigerung, da sie die Schaltvorgänge der Leistungshalbleiter präziser und dynamischer steuern können. In Kombination mit Batterie-Inverter-Systemen wie Modular Multilevel Converters (M2LC) können Systemeffizienz und Modularität erheblich verbessert werden \cite{RiarICIT2013}. Die Verwendung von WBG-Halbleitern, etwa SiC oder GaN, in diesen Systemen bietet zusätzliche Vorteile, insbesondere durch die Reduktion von Verlusten und eine höhere Betriebstemperaturfestigkeit. Der Einsatz solcher Technologien, kombiniert mit der Optimierung der Systemarchitektur und innovativen Ansätzen wie der Schalterhybridisierung, führt zu einem zukunftsfähigen und hocheffizienten Antriebsstrang \cite{SachsTrucks2025}. Tabelle 1 stellt diese und weitere Faktoren für eine Optimierung der Systemeffizienz dar.

\begin{table}[htbp]
\centering
\footnotesize
\setlength{\tabcolsep}{4pt}
\renewcommand{\arraystretch}{1.05}
\begin{tabularx}{\textwidth}{@{} L{0.34\textwidth} Y Y @{}}
\toprule
\textbf{Optimierungsstrategie} & \textbf{Ermöglicht die Reduktion von} & \textbf{Trade-off} \\
\midrule
WBG-Halbleiter (z.\,B.\ SiC, GaN) {\cite{shi2023_ev_inverter_review}} & Schaltverluste, Leitungsverluste & Kosten, Motorlebensdauer, EMC \\
Reduzierte Streuinduktivität {\cite{sun2021_power_loop,stewart2016_dc_bus}} & Schaltverluste & Mechanische Komplexität \\
Variable Schaltgeschwindigkeit {\cite{ti2023_ucc5880,ivanis2024_sic_gatedrive}} & Schaltverluste & Kosten, Motorlebensdauer, EMC \\
Anpassung der Schaltfrequenz {\cite{kim2024_pwm_nvh,zhang2024_pmsm_noise_review}} & Schaltverluste, harmonische Verluste & NVH, Spannungsripple, EMC \\
Schalterhybridisierung {\cite{pai2018_sic_hybrid,sheikhan2024_hybrid_switch,delft2022_hybrid_gate}} & Schaltverluste & Zusätzliche Gatetreiber-/Elektronik, Varianten \\
Smarte Gatetreiber {\cite{ti2023_ucc5880,ivanis2024_sic_gatedrive}} & Schaltverluste & Komplexität \\
Hairpin-Wicklungen {\cite{shams2022_hairpin,pastura2022_hairpin_end}} & Kupferverluste (DC/AC) & Hochgeschwindigkeitseffizienz \\
Reduzierte Blechdicke {\cite{xia2020_joining,jfe2006_motorloss}} & Wirbelstromverluste, harmonische Verluste & Kosten, mech. Stabilität \\
Geklebte Blechpakete {\cite{xia2020_joining}} & Wirbelstromverluste & Kosten \\
Reduzierte Polzahl {\cite{jfe2006_motorloss}} & Eisenverluste, harmonische Verluste & Leistungseinbußen \\
Magnetsegmentierung {\cite{energies2019_pm_losses,zhang2021_iet_segmentation}} & Harmonische Verluste (Magnet) & Kosten \\
Trennungseinheit (AWD) {\cite{xu2023_adaptive_disconnect,nhtsa2019_awd_disconnect}} & Reibungsverluste & Kosten \\
Mehrmotor-Konfigurationen {\cite{wu2018_dualmotor,ganesan2023_mi_mpc}} & Systemverluste & Komplexität, Kosten \\
Variable Schaltfrequenz {\cite{ibrahim2024_vsfhpwm}} & Schaltverluste, harmonische Verluste & NVH, EMC, DC-Ripple \\
Alternative Modulationsmethoden {\cite{hava1998_gdpwm,kim2024_pwm_nvh}} & Schaltverluste, harmonische Verluste & NVH, EMC, DC-Ripple \\
Optimierte Motoransteuerung (MTPE, MTPA, MTPV) {\cite{dianov2022_mtpa,ti2024_motor_control}} & Eisen-, Kupfer-, Inverterverluste & Komplexität \\
Thermische Feldschwächung & Eisen-, Kupfer-, Inverterverluste & Inverterlebensdauer \\
Holistische Softwarestrategien & Systemverluste & Komplexität \\
Optimierung der Zwischenkreiskondensatoren {\cite{sachsASIA2025,holm2018_sic_dclink,liu2021_dclink_soa}} & Systemverluste, Kosten & Thermische \& mech. Stabilität \\
Modular Multilevel Converters  & Systemverluste & Modularitätskosten \\
Holistische Systemarchitektur & Systemverluste & Entwicklungsaufwand \\
\bottomrule
\end{tabularx}
\end{table}

%% file: conclusion.tex
\section{Zusammenfassung}

Die vorgestellten Konzepte und Technologien zur Effizienzsteigerung von Traktionsmotoren in batterieelektrischen Fahrzeugen zeigen, dass eine signifikante Reduktion der Verluste im Antriebsstrang möglich ist. Dabei erweisen sich teillastoptimierbare Inverter- und Motortopologien als besonders vielversprechend, da sie durch innovative Ansätze wie Mehrlevel-Inverter, dynamische Spannungsanpassungen und flexible Ansteuerstrategien eine optimierte Balance zwischen Effizienz und Kosten erreichen können.

Die Auswahl der geeigneten Technologie hängt von mehreren Faktoren ab, darunter Effizienz, Kosten, Leistungsdichte sowie der Umwelteinfluss der verwendeten Materialien und Prozesse. Neuartige Halbleitermaterialien wie SiC ermöglichen deutliche Effizienzgewinne, während eine durchdachte Systemarchitektur und innovative Steuerungskonzepte die Gesamtsystemleistung weiter steigern können. Insgesamt zeigt sich, dass keine einzelne Lösung alle Anforderungen gleichermaßen erfüllen kann. Stattdessen ist eine Kombination aus verschiedenen Ansätzen erforderlich, die je nach Anwendungsszenario individuell abgestimmt werden müssen. Entscheidend für die Durchsetzung neuer Konzepte ist ihre Fähigkeit, technologische Effizienz mit ökonomischer Machbarkeit und Nachhaltigkeit zu verbinden. Nur so können zukunftsfähige Antriebssysteme realisiert werden, die den Anforderungen moderner Elektromobilität gerecht werden.

%% file: 99_acronymsA.tex

\acrodef{FEA}{finite element analysis}
  \acrodefplural{FEA}[FEAs]{finite element analyses}
  
\acrodef{BEV}{battery electric vehicle}
  \acrodefplural{BEV}[BEVs]{Battery electric vehicles} %

\acrodef{BMS}{battery‑management system}%
  \acrodefplural{BMS}[BMSs]{battery‑management systems}%

\acrodef{WLTP}{worldwide harmonized light vehicles test procedure}%

\acrodef{OEM}{original equipment manufacturer}%
  \acrodefplural{OEM}[OEMs]{original equipment manufacturers}%
\acrodef{iPMSM}{interior permanent‑magnet synchronous motor}%
  \acrodefplural{iPMSM}[iPMSMs]{interior permanent‑magnet synchronous motors}%
  \acrodefplural{LUT}[LUTs]{look-up tables}%


\acrodef{WBG}{wide-bandgap}
\acrodef{GaN}{Gallium Nitride}
\acrodef{SiC}{silicon carbide}
\acrodef{Si}{silicon}
\acrodef{IGBT}{insulated‑gate bipolar transistor}%
\acrodef{MOSFET}{metal‑oxide semiconductor field‑effect transistor}%


\acrodef{VSI}{voltage-source inverter}
  \acrodefplural{VSI}[VSIs]{voltage-source inverters}%

\acrodef{MLI}{multi-level inverter}
\acrodef{2L}{two-level}
\acrodef{3L}{three-level}
\acrodef{5L}{five-level}
\acrodef{Q5L}{quasi-five-level}
\acrodef{SM}{switching matrix}

\acrodef{DM}{differential-mode}
\acrodef{CM}{common-mode}

\acrodef{3‑Φ}{three-phase}

\acrodef{FC}{flying capacitor}
\acrodef{TNPC}{T-type neutral point clamped}
\acrodef{ANPC}{active neutral point clamped}
\acrodef{B6}{six halfbridge}


\acrodef{PWM}{pulse width modulation}%
\acrodef{RMS}{root‑mean‑square}%
\acrodef{SVPWM}{space-vector pulse-width modulation}%


\acrodef{AWD}{all‑wheel drive}%
  \acrodefplural{AWD}[AWDs]{all‑wheel drives}
\acrodef{FWD}{front‑wheel drive}%
  \acrodefplural{FWD}[FWDs]{front‑wheel drives}%
\acrodef{RWD}{rear‑wheel drive}%
  \acrodefplural{RWD}[RWDs]{rear‑wheel drives}%

\acrodef{BMS}{battery‑management system}%
\acrodef{DSC}{double‑sided‑cooled}%
\acrodef{OBC}{on‑board charger}%
\acrodef{PDU}{power‑distribution unit}%